\documentclass{eas}
\usepackage{graphicx}
\newcommand{\Teff}{$T_{\rm eff }$}
\TitreGlobal{Setting a new standard in the analysis of binary stars}

\begin{document}

\title{Probing the models: Abundances for high-mass stars in binaries}

\runningtitle{Pavlovski \& Southworth: Abundances for high-mass stars in binaries}

\author{K.\ Pavlovski}\address{Department of Physics, Faculty of Science,
          University of Zagreb, 10\,000 Zagreb, Croatia}
\author{J.\ Southworth}\address{Astrophysics Group, Keele University, Staffordshire
             ST5 5BG, UK}

\begin{abstract}
The complexity of composite spectra of close binary star system makes study of
the spectra of their component stars extremely difficult. For this reason there
exists very little information on the photospheric chemical composition of stars
in close binaries, despite its importance for informing our understanding of the
evolutionary processes of stars. In a long-term observational project we aim to
fill this gap with systematic abundance studies for the variety of binary systems.
The core of our analysis is the spectral disentangling technique, which allows
isolation of the individual component star spectra from the time-series of
observed spectra.

We present new results for high-mass stars in close binaries. So far, we have
measured detailed abundances for 22 stars in a dozen detached binary systems.
The parameter space for the stars in our sample comprises masses in the range
8--22 M$_\odot$, surface gravities of 3.1--4.2 (c.g.s.) and projected rotational
velocities of 30--240\,km$\,$s$^{-1}$. Whilst recent evolutionary models for
rotating single stars predict changes in photospheric abundances even during
the main sequence lifetime, no star in our sample shows signs of these predicted
changes. It is clear that other effects prevail in the chemical evolution of
components in binary stars even at the beginning of their evolution.
\end{abstract}

\maketitle

%%%%%%%%%%%%%%%%%%%%%%%%%%%%%%%%%%%%%%%%%%%%%%%%%%%%%%%%%%%%%%%%%%%%%%%%%%%

\section{Motivation}

Detached eclipsing binary stars (dEBs) are the primary source of fundamental
stellar quantities: mass $M$ and radius $R$. Both components must be
spectroscopically detectable for direct determination of these quantities.
Modern astronomical instrumentation and analytical methods enable a high
accuracy and precision to be reached, reducing errors in the masses and
radii for the stars in binaries to 1\% or less (e.g.\ Pilecki \etal\ 2013).
A critical examination of published analyses found 95 systems satisfying
the criterion of 3\% in the uncertainties for $M$ and $R$ (Torres \etal\ 2010).
The DEBCat catalogue lists 159
systems\footnote{\tt http://www.astro.keele.ac.uk/jkt/debcat/}.

For a useful comparison with theoretical evolutionary models further observables
are needed: effective temperature, \Teff, and metallicity, [M/H]. Only the ratio
of the \Teff s of the two stars is provided by the light curve analysis. The
\Teff\ of one component must be determined from other sources, e.g.\ spectral
energy distribution, colour indices, or spectral type estimates. Far worse is
the situation with empirical determination of the components' metallicity. Of
the 95 systems listed by Torres \etal\ (2010), fewer than half have an empirically
estimated metallicity. Detailed abundance studies has been performed for only
four systems, including two from our own work (see Sect.~5.).

The method of Doppler tomography (Bagnuolo \& Gies 1991), and its generalisation
as spectral disentangling (Simon \& Sturm 1994, Hadrava 1995) enables the study
of the individual spectra of the components of binary star systems. This is
particularly useful for the determination of \Teff\ and metallicity (c.f.\
Pavlovski \& Southworth 2012, and references therein). Since the structure
and evolution of a star depends on its chemical composition, the degeneracy
in the location of a star in evolutionary diagrams can only be broken if
an empirical metallicity is known.

\section{Theoretical framework}

In the last two decades a new generation of stellar evolutionary models has
been developed (c.f.\ Maeder \& Meynet 2012, Langer 2012). The inclusion of
stellar rotation has had profound effects on these models (Meynet \& Maeder 2000,
Heger \& Langer 2000, Heger \etal\ 2000). Centrifugal force changes the stellar
shape and structure, causes meridional circulation and induces turbulent mixing.
In turn, the \Teff\ and luminosity of the rotating star changes, affecting its
liftime. As a consequence, the surface abundance pattern changes too, heavily
affecting the expected photospheric nitrogen abundance.

The calculations predict an increase in nitrogen abundance with increasing
stellar mass, increasing initial rotational velocity, and decreasing metallicity.
This makes nitrogen abundance viable as a new observable (K\"{o}hler \etal\ 2012).
However, caution is needed since Hunter \etal\ (2009) have found from the
observations of a large sample of the Galactic and Magellanic Clouds B stars
the relation between nitrogen enhancement and stellar parameters (mass and
projected rotational velocity, $v\sin i$) to be more complex than predicted
from the models. Subsequently, Morel \etal\ (2008) did not find any tight
correlation of the nitrogen abundance with the strength of the magnetic field.
This was also supported by extension of the analysis to O stars (Martins \etal\ 2012).

Recently, comprehensive grids of theoretical stellar evolutionary models incorporating
 rotation have been available by the Utrecht (Brott \etal\ 2011) and Geneva
(Ekstr\"{o}m \etal\ 2012) groups. These models differ in the treatment of
rotational mixing, as well as in the calibration. Brott \etal\ (2011) tailored
their grids of different initial composition to match the results of {\sc vlt-flames}
survey of massive stars in the Magellanic Clouds and the Galaxy (Evans \etal\ 2005).
In consequence, their models have no scaled-solar composition, and result in a rather
low metallicity for Galactic stars, Z$ = 0.088$. In contrast, Ekstr\"{o}m \etal\
(2012) use scaled-solar compositions, and galactic models are represented with the
metallicity Z$ = 0.014$.

\section{Spectral disentangling}

The complexity of composite spectra of close binaries makes the study of the
individual stellar spectra extremely difficult. Shifts of the spectral lines due
to the orbital motion of components in binary or multiple systems are essential
for the determination of stellar masses. Spectral lines overlap for much of each
orbital cycle, and the secondary star may be much fainter than the primary, and
thus contribute only a small fraction of the total light of the system. To these
problems should be added the intrinsic broadening of spectral lines in OB stars,
and high $v\sin i$ in binaries. All these pose a difficulty for accurate
measurements of radial velocities (RVs) for the components, lowering the
accuracy of the derived stellar masses (c.f.\ Andersen 1991, Torres \etal\ 2010)
or making the fainter star impossible to detect. Since the first spectroscopic
detection of binary stars in the last decades of the 19th century, binary star
spectra have been used almost solely for RV measurement. Important astrophysical
information contained in these spectra was ignored due to the inability of
researchers to extract individual spectra of the components.

The method of {\em spectral disentangling} ({\sc spd}) is the way to avoid these
obstacles. Bagnuolo \& Gies (1992) successfully applied the method of Doppler
tomography and isolated component star spectra for the case when the RVs of
the individual exposures were already known. Simon \& Sturm (1994) generalised
tomographic separation by setting a matrix equation from a time-series of binary
star spectra. A set of linear equations is solved directly for the orbital elements
 and the component spectra, thus bypassing the determination of RVs entirely.
An independent formulation of the equations of {\sc spd} in the Fourier space
by Hadrava (1995) has been particularly advantageous since the calculations
are more efficient and computationally less demanding. An overview of {\sc spd}
and its variants is given by Pavlovski \& Hensberge (2010).

\begin{figure}[h]
\begin{center}
\includegraphics[width=0.74\textwidth]{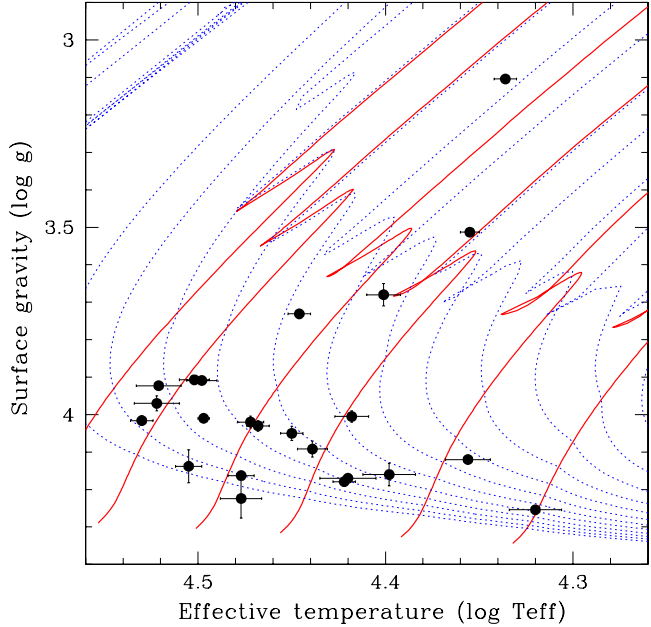}
\end{center}
\caption{Location of the stars in our sample of eclipsing
binary systems versus the theoretical evolutionary tracks
and isochrones from Ekstr\"{o}m \etal\ (2011).}
\end{figure}

In comprehensive studies of V578\,Mon, Hensberge \etal\ (2000) and Pavlov\-ski
\& Hensberge (2005) constructed a complete procedure for the analysis of
spectroscopically double-lined eclipsing binary stars. In the core of the
procedure is {\sc spd}, which yields the optimal set of orbital elements
and isolated individual spectra of the components. Further, the analysis
of these spectra provides the atmospheric parameters for the components:
\Teff s, metallicity, detailed abundances and $v\sin i$ values. In turn,
the atmospheric parameters are fed back into fine-tuning the light curve
solution. In an iterative cycle an optimal set of stellar and binary
parameters can be achieved. Further development and applications are given
in Pavlovski \& Southworth (2009) and Pavlovski et \etal\ (2009). This scheme
makes possible the determination of the fundamental quantities -- $M$, $R$,
\Teff\ -- with the accuracy needed for a useful comparison to theoretical models.

Complementary observables hold an important possibility in the analysis of
binary and multiple stars. Only a complementary solution of astrometric and
spectroscopic measurements could yield a complete set of the orbital elements
and its space orientation. With the recent tremendeous development in
interferometric capability, many more spectroscopic binaries have been
spatially resolved. The stability and quality of the solution in {\sc spd}
could be supported with constraints from external observables such as
interferometric measurements (Kolbas \& Pavlovski, this proceedings).

An important step in the analysis is determination of the \Teff s for
the stars in binary systems, directly from renormalised disentangled spectra.
The traditional degeneracy between \Teff\ and surface gravity can be lifted
for the stars in binary systems. The surface gravities for these star can be
determined with uncertainties below 0.01\,dex.

A promising and important new development is modelling of the composite 
spectra for tidally and rotationally distorted stars in binary system given 
in Palate \etal\ (2013).

\begin{figure}[h!]
\begin{tabular}{cc}
\includegraphics[width=0.49\textwidth]{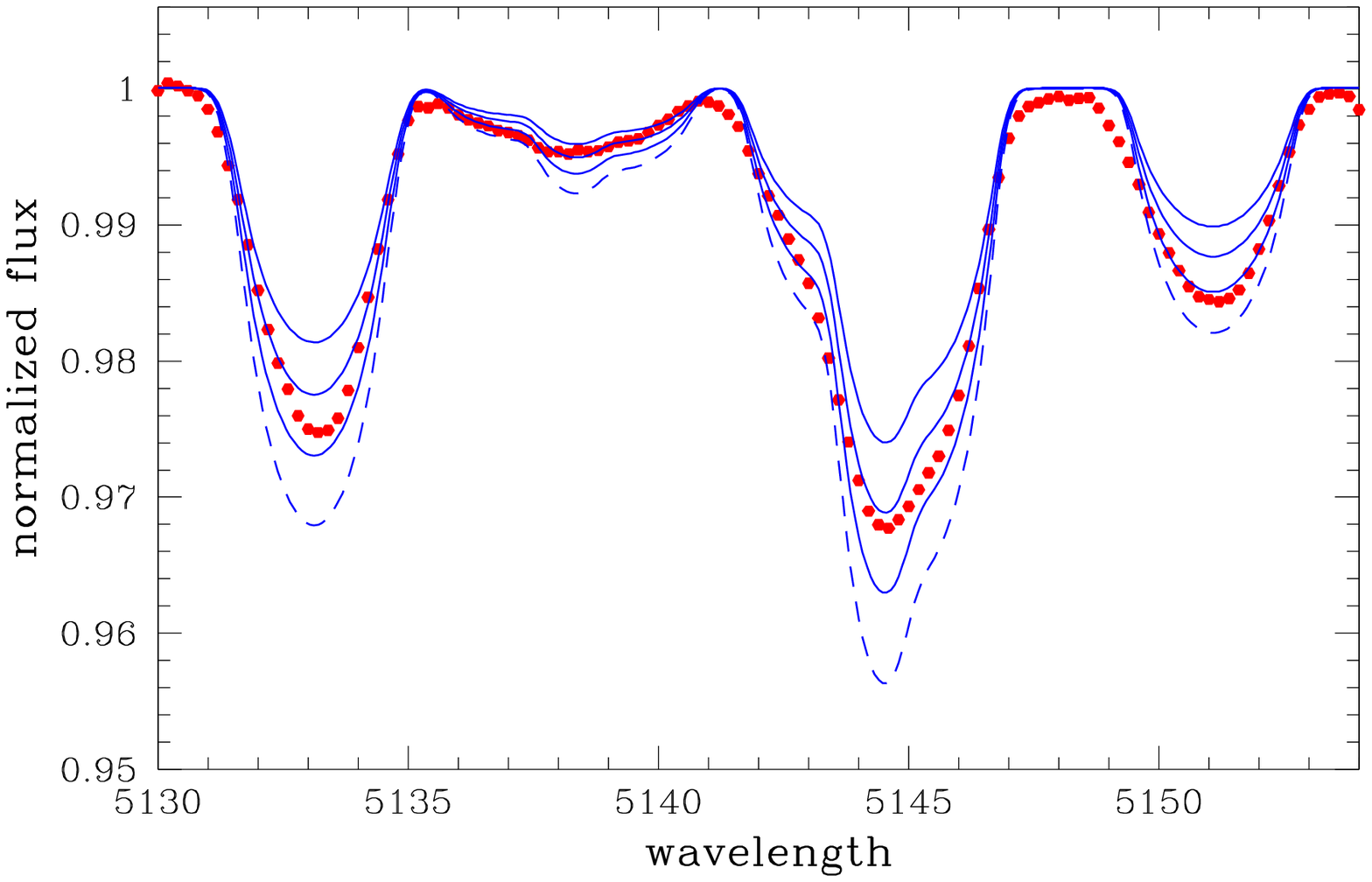}
& \includegraphics[width=0.49\textwidth]{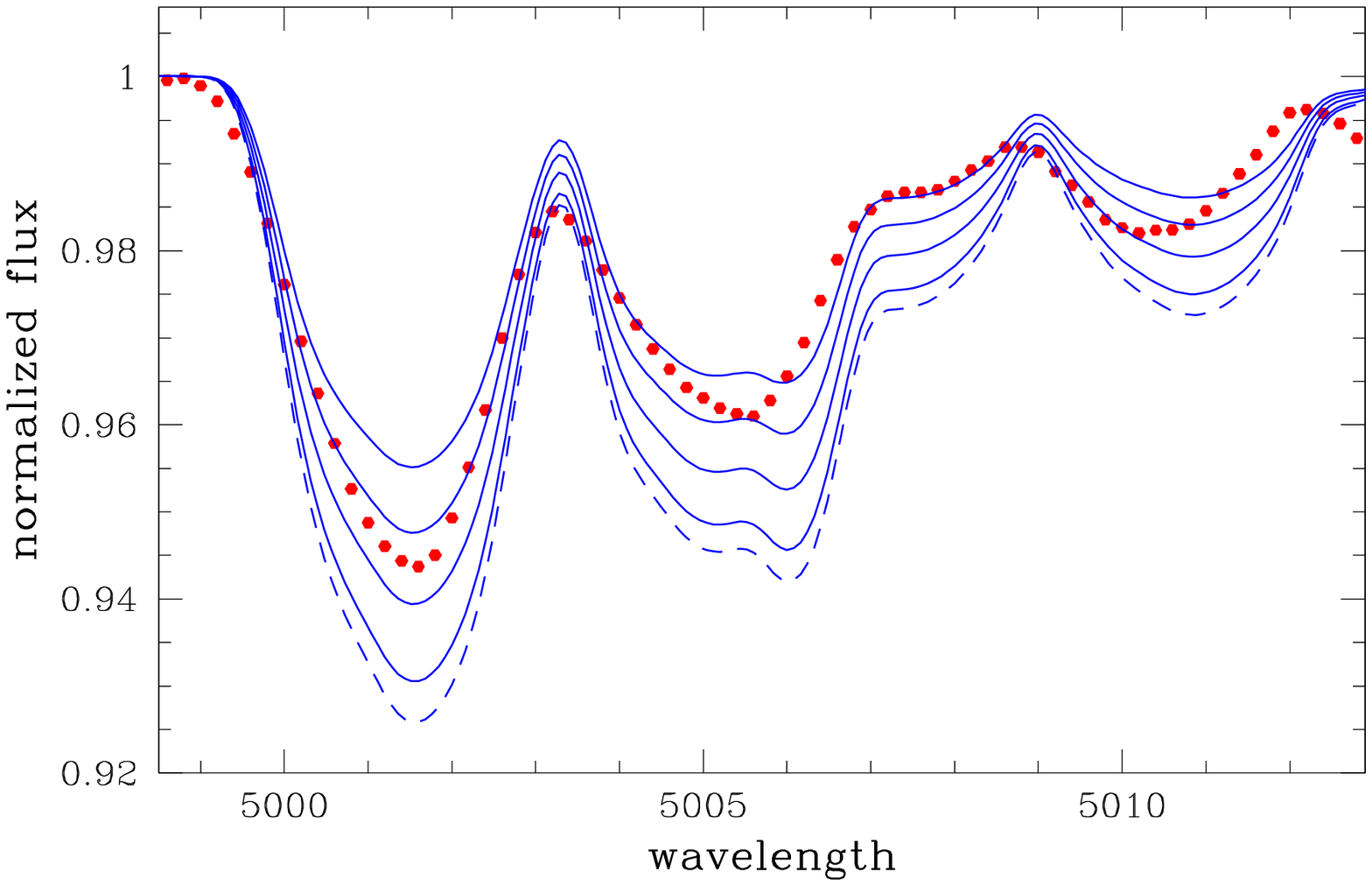}
\end{tabular}
\caption{Comparison between renormalised disentangled spectrum (symbols)
of the giant component in V380\,Cyg with the synthetic spectra (lines)
computed assuming different abundances of C (left panel) and N (right
panel). Dashed lines correspond to the 'present-day abundances' as
reported by Nieva \& Przybilla (2012).
\label{v380abund}}
\end{figure}

\section{Abundances for high-mass stars in binary systems}

So far, we have observed and analysed a dozen binary stars with high-mass
components. Their location in the Hertzsprung-Russell diagram (HRD) is
shown in Fig.\,1 along with theoretical evolutionary tracks and isochrones.
The majority of the stars are in early phases of their main-sequence evolution,
with the youngest systems being V573\,Car and V578\,Mon. The stars in our sample
are late-O to early-B, i.e.\ at the lower end of the high-mass stars. The largest
masses are 20--25\,M$_\odot$ for the components of V1034\,Sco, DN\,Cas and HD\,164258.
Their $v \sin i$ values range from $\sim$30\,km\,s$^{-1}$ (V621\,Per\,A;
Southworth \etal\, in preparation) to 250\,km\,s$^{-1}$ (HD\,164258\,A;
Mayer \etal\ 2013). Of the systems studied, V380\,Cyg, V453\,Cyg and V621\,Per
are the  most informative as their primary components are evolved to either
close to or beyond the terminal-age main sequence (TAMS). Here we described
some principal results and summarise published results for the most important
systems.\\[-5pt]

\begin{figure}[h!]
\begin{center}
\includegraphics[width=0.62\textwidth]{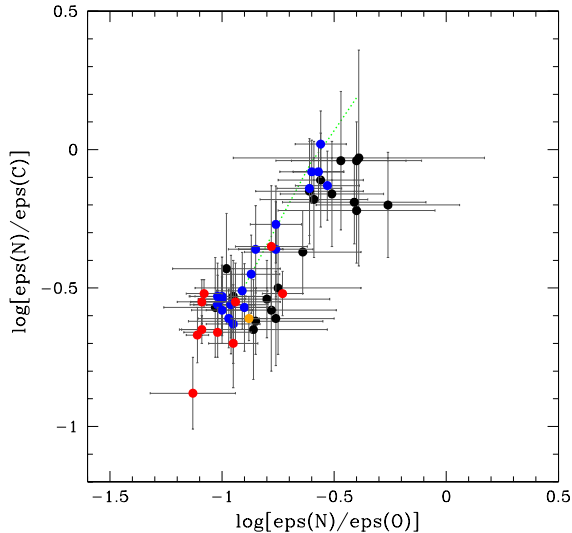}
\end{center}
\caption{The trend in N/C versus N/O ratios found for B stars. Three
samples of B stars are shown: single stars from Nieva \& Przybilla (2012)
(blue circles), magnetic stars from Morel \etal\ (2008) (black circles)
and stars in close binaries (this work, red circles).}
\end{figure}

{\bf V578\,Mon} (HD\,259135, BD\,+4$^\circ$1299) is an eccentric dEB, in the
young open cluster NGC\,2244 embedded in the Rosette Nebula, with orbital
period 2.408\,d. Hensberge \etal\/ (2000) determined the absolute dimensions
of the components from $uvby$ photometry and high-resolution spectroscopy.
They found the binary to consist of B1\,V + B2\,V stars with masses 14.5
and 10.3\,M$_\odot$, radii 5.2 and 4.3\,R$_\odot$, and $v \sin i$ values
114 and 98\,km\,s$^{-1}$. The age inferred from isochrone fitting was
estimated to be $\sim$3\,Myr. Garcia \etal\ (2011, 2013) provided the first
measurement of the apsidal motion period and improved the orbital and stellar
parameters.
A new comprehensive analysis of all available photometric data plus a new
\'{e}chelle spectroscopy with Mercator/{\sc hermes}, has allowed a fine-tuning
of the physical properties and the internal structure of the stars
(Garcia \etal\ 2014). The photospheric chemical composition of the components
has been determined using disentangled spectra (Pavlovski \& Hensberge 2005).
Their analysis was done relative to sharp-lined B1\,V star \#201 in NGC\,2244
for which accurate abundances are known (Vrancken \etal\ 1997). The abundance
pattern for both stars is consistent with those found for several B1 stars
in NGC\,2244 (Vrancken \etal\ 1997).\\[-5pt]

{\bf V453\,Cyg} (HD\,227696) is totally eclipsing dEB for which
accurate physical properties have been measured (Southworth \etal\
2004a). The system consists of two B stars with masses 14.4 and
11.1\,M$_\odot$. The more massive component is approaching the TAMS
whilst its companion has completed about half of its main-sequence
lifetime. These characteristics make V453\,Cyg ideal for probing the
main-sequence chemical evolution of high-mass stars. The
photospheric chemical composition of the components was determined
by Pavlovski \& Southworth (2009). The more massive component is the
only star in our sample which shows the trend in CNO abundances
expected on theoretical grounds (Sect.\,2, see also Przybilla, this
volume). In Fig.\,3 the observational results are presented for the
three samples: single B stars (Nieva \& Przybilla 2012), single
magnetic B stars (Morel \etal\ 2008) and B stars in close binaries
(this work). The trend in the changes of abundance ratio between N/O
and N/C is governed by the CNO cycle and stellar evolution
(Przybilla \etal\ 2010;
Przybilla, this volume).\\[-5pt]

{\bf V380\,Cyg} (HD\,187879, KIC\,5385723) is an eccentric dEB with
an evolved B-type component ($\log g = 3.1$). A large discrepancy between
the dynamical and evolutionary masses for this component has been found.
Guinan \etal\ (2000) removed this mass discrepancy by invoking extremely
strong convective core overshooting. Pavlovski \etal\ (2009) tried
unsuccesfully to match the giant component to the rotating evolutionary
models of Ekstr\"{o}m \etal\ (2008). As a potential candidate for stochastic
pulsations, concerted effort has been put in to observe the binary
photometrically with the {\em Kepler} satellite, and spectroscopically
with Mercator/{\sc hermes} (Tkachenko \etal\ 2012, 2013). This has allowed
revision of the absolute dimensions and atmospheric diagnostics for both
components even though the secondary contributes only about 6\% of the
total light of the system. The measured chemical composition for the
giant component corroborates our previous study (Pavlovski \etal\ 2009).
Spectral disentangling of a large collection of high-resolution spectra
also makes possible a study of the secondary component. Its composition
agrees with the primary's to within the uncertainties. N/C and N/O are
characteristic for unevolved B stars, despite one of the component having
 evolved to the giant stage.\\[-5pt]

{\bf V621\,Per} (BD\,+56$^\circ$576) is a totally eclipsing binary system,
and a member of the young open cluster $\chi$\,Per (NGC\,884). The primary
star is an early-type subgiant (Southworth \etal\ 2004b). In an extensive
photometric campaign on $\chi$\,Per, complete multicolour light curves
of this long-period ($P = 25.5$ d) binary have been obtained (Saesen \etal\
2010). An application of spectral disentangling reveals the secondary
component which makes possible the determination of the absolute dimensions
of the components (Southworth \etal\, in preparation). It is found that
the subgiant component has $M = 11.2$\,M$_\odot$, and $\log g = 3.5$.
Its helpfully low rotation rate ($v\sin i = 32$\,km\,s$^{-1}$), allows
an accurate determination of its atmospheric parameters and chemical
composition. We find that the nitrogen and carbon abundances do not
show the theoretically predicted excess, thus corroborating our results
for the giant component of V380\,Cyg.

\section{Conclusion}

We have determined the photospheric abundances for helium and CNO elements
in high-mass stars in close binaries. We do not find any dependence on
their masses, rotational velocities or evolutionary status except for
the marginal case of V453\,Cyg\,A. The obvious question is: are high-mass
 stars in detached binary systems fundamentally different to single high-mass
stars for which changes in the photospheric abundances have been disclosed
observationally (Hunter \etal\ 2009, Przybilla \etal\ 2010, see also Przybilla
this volume)? Obviously, tidal effects have some effect on the stellar
structure and evolution, and may hinder rotationally induced mixing in
high-mass components in close binary systems. This was calculated and
discussed by de Mink \etal\ (2009, 2013). However, a significantly larger sample
is needed before firm conclusions can be drawn.

%%-----------------------------
%%      your bibliography
%%-----------------------------

\end{document}